# Search for relativistic Magnetic Monopoles with ten years of the ANTARES detector data

J. Boumaaza,[a] J. Brunner,[b] A. Moussa,[c] and Y. Tayalati,[a] on behalf of the ANTARES Collaboration

[a] *University Mohamed V in Rabat, Faculty of Science*
 *4 Avenue Ibn Battouta B.P. 1014 RP, Rabat, Morocco*

[b] *Centre de Physique des Particules de Marseille, France*

[c] *University Mohammed First in Oujda, Morocco*
 E-mail: boumaaza.jihad@gmail.com

ABSTRACT: The presented study is an updated search for magnetic monopoles using data taken with the ANTARES neutrino telescope over a period of 10 years (January 2008 to December 2017). In accordance with some grand unification theories, magnetic monopoles were created during the phase of symmetry breaking in the early Universe, and accelerated by galactic magnetic fields. As a consequence of their high energy, they could cross the Earth and emit a significant signal in a Cherenkov-based telescope like ANTARES, for appropriate mass and velocity ranges. This analysis uses a run-by-run simulation strategy, as well as a new simulation of magnetic monopoles taking into account the Kasama, Yang and Goldhaber model for their cross section with matter. The results obtained for relativistic magnetic monopoles with $\beta = v/c \geq 0.817$, where v is the magnetic monopole velocity and c the speed of light in vacuum, are presented.

# Contents

# 1. Introduction

The existence of magnetic charges has been considered long ago. The introduction of hypothetical magnetic charges and magnetic currents can restore the symmetry in the Maxwell's equations with respect to magnetic and electric fields.

When investigating the symmetry between electricity and magnetism, Paul Dirac proved in 1931 [1], that the introduction of Magnetic Monopoles (MMs) can also elegantly solve the mystery of the quantization of electric charge. In addition to this, Grand Unified Theories (GUTs) also predicted that MMs could be created shortly after the Big Bang.

In a Cherenkov based telescope such as ANTARES (Astronomy with a Neutrino Telescope and Abyss environmental RESearch) [2], the signal of a MM would be distinguishable from atmospheric muons and neutrinos due to the importance of the amount of the light emitted by MMs.

Several searches were carried out using neutrino telescopes. The ANTARES neutrino telescope results of the analyses published in [3] and [4] using data sets of 116 days and 1012 days respectively, as well as the results of the IceCube collaboration [5] can be seen in Figure 5.

This work comes as a sequel to previous analyses, taking into consideration a higher statistic (2480 days) and a change in the model for the MMs cross section with matter.

# 2. Magnetic Monopoles

Magnetic Monopoles (MMs) are particles with one magnetic pole, and are supposed to be the magnetic counterparts of electric charges (electrons). MMs are topologically stable particles and carry a magnetic charge defined as a multiple integer of the Dirac charge:

$$g_D = \frac{\bar{h}c}{2e} = \frac{e}{2\alpha} = 68.5 \, e. \quad \quad (2.1)$$

where $e$ is the electric charge, $c$ is the speed of light in vacuum, $\bar{h}$ is the Planck constant and $\alpha \simeq 1/137$ is the fine structure constant.

While Dirac had demonstrated the consistency of MMs with quantum electrodynamics, 't Hooft [6] and Polyakov [7] proved the necessity of MMs in Grand Unification Theories (GUT). This led to the conclusion that any unification model in which the U (1) subgroup of electromagnetism is embedded in a semi-simple gauge group and which is spontaneously broken by the Higgs mechanism possesses monopole-like solutions. The masses of MMs can range from $10^8$ to $10^{17}$ GeV/c$^2$.

Moreover, MM would be created after the Big Bang (during the phase transition of symmetry breaking), and they would be accelerated by galactic magnetic fields if their mass was under $10^{14}$ GeV/c$^2$. The rarity of GUT MMs is also a motivation to the scenario of inflation.

## 3. Event selection

In order to remove the background from atmospheric muons and neutrinos, a number of cuts is used. The first one is applied on the Zenith angle ≥ 90° and it aims to select only upgoing events, using the Earth as a filter, and eliminating all the atmospheric down going background. In the second one, we require only events reconstructed with at least 2 lines of the detector (*nlines* ≥ 2) to further improve the quality of the events. And finally, to prioritize the selection of events with a fit quality parameter rather than a bright point one, the last primary cut chosen is that the quality of reconstructed tracks parameter (tχ2) must be inferior to the quality of reconstructed bright points (showers) parameter (bχ2). The Nhit parameter defined earlier is chosen as a discriminant variable in this study, since it can refer to the amount of light emitted by the particle, and knowing that MMs are expected to produce large amounts of light compared to other particles, a cut on Nhit is seen a powerful tool to distinguish the MMs signal from the background. An other discriminant variable is used to isolate the MM signal from the background is α, which is the reconstruction quality taking into account the brightness of the event.

$$\alpha = \frac{t\chi 2}{1.3 + (0.04 \times (\text{Nhit} - N_{df}))^2} \quad . \tag{3.1}$$

$N_{df}$ indicates the number of the free parameters reconstructed. It is equal to 5 in this case. Figure 2 shows a comparison between data sample and MC for the Nhit and α distributions, a good agreement can be observed.

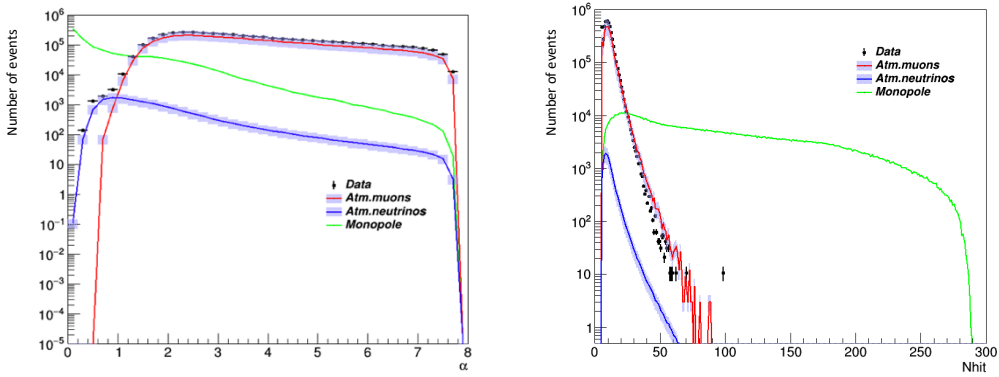

**Figure 1.** Nhit distribution on the right and α distribution on the left showing the signal of MM simulated in the interval of β in [0.86, 0.90] as an example, (green histogram), the background of atmospheric muon (red) and neutrinos (blue) and real data (Sample of 10% in black dots).

.

## 4. Optimization

This section aims to optimize the cuts on Nhit and α to be applied in order to isolate the MMs signal from the background. Figure 2 below shows the event distribution of the scatter plot α versus Nhit for the range [0.86, 0.90] of β as an example. The MMs signal can be distinguished from the background by applying a couple of cuts on α and Nhit. For slow MMs, an extra cut on β reconstructed allows further suppression of the background.

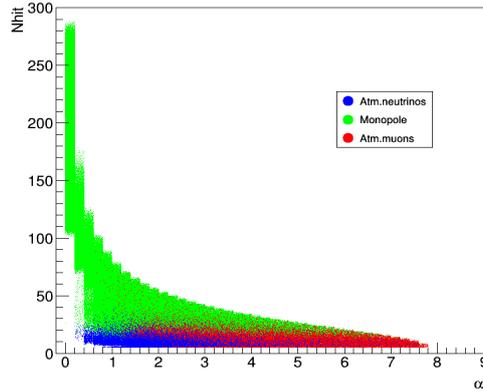

**Figure 2.** Scatter-plot of the two variables α and Nhit, the MMs signal in green simulated with β in the range [0.86, 0.90] and the background regions consisting of atmospheric muons in red and atmospheric neutrinos in blue.

### 4.1 Extrapolation of the atmospheric muon distribution

In order to compensate for the lack of statistics in the Nhit distribution for muons an extrapolation has been made in the region, by fitting the histogram with a Landau distribution as seen in figure 3. The total background events used to calculate the sensitivity include the contribution of extrapolation. This method of extrapolation allows the recovery of the statistical errors of the simulation.

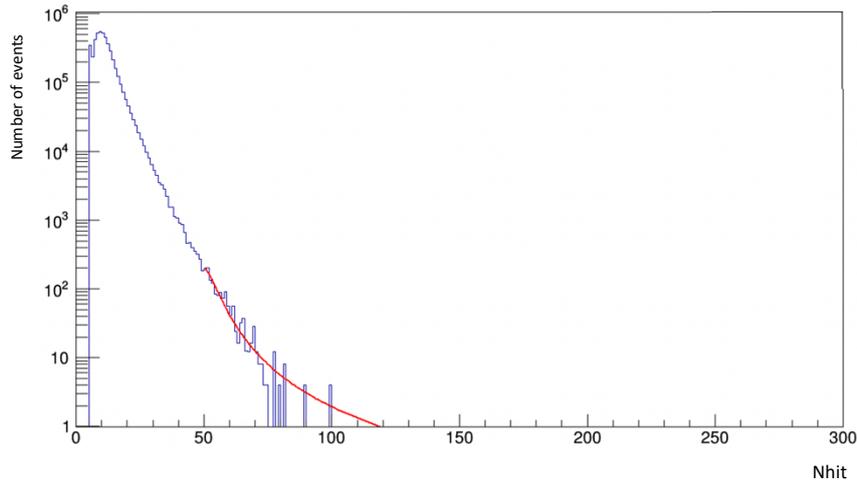

**Figure 3.** Distribution of Nhit for atmospheric muons representing the extrapolation after applying a fit using a Landau type function in red, the extrapolation is taken into account when calculation the sensitivity.

## 4.2 Model Rejection Factor

To obtain the best sensitivity we optimize the Model Rejection Factor [8] for each velocity interval, by relying on α and Nhit cuts. The 90% C.L. sensitivity $S_{90\%}$ is calculated using the Feldman-Cousins [9] formula, considering events which follow a Poissonnian distribution.

$$S_{90\%}(cm^{-2}\ s^{-1}\ sr^{-1}) = \frac{\bar{\mu}_{90}(n_b)}{S_{eff}(cm^2.sr) \times T_{(s)}} \quad , \quad (4.1)$$

where T is the duration of data taking, and $\bar{\mu}_{90}$ and $S_{eff}$ are defined as:

$$\bar{\mu}_{90}(n_b) = \sum_{n_{obs}=1}^{\infty} \mu_{90}(n_{obs}, n_b) \frac{n_b^{n_{obs}}}{n_{obs}!} e^{-n_b} \quad , \quad (4.2)$$

$$S_{eff} = \frac{n_{MM}}{\Phi_{MM}} \quad . \quad (4.3)$$

with $n_{MM}$ is the number of MMs remaining after cuts, and $\Phi_{MM}$ represents the flux of generated MMs. The Model Rejection Factor consists of varying the cuts until the minimum of Rejection Factor (RF) is found, which coincides with the best sensitivity. After the optimization of the rejection factor RF, sensitivity at 90% C.L. is derived using the Feldman-Cousins formula.

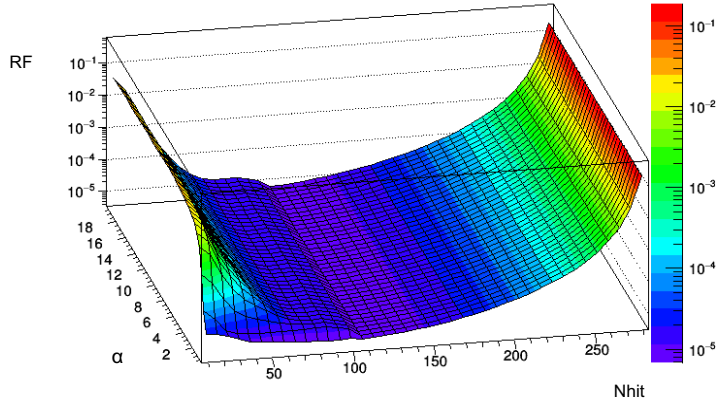

**Figure 4.** The model rejection factor RF as a function of $\alpha$ and Nhit cuts represented here for $\beta$ in the range [0.86, 0.90].

## 5. Results and conclusion

The sensitivities obtained corresponding to 10 years of data, as well as the cuts and the number of events remaining for each speed. are summarized in the table below:

| β interval | α cut | Nhit cut | Background events | Sensitivity (cm⁻² s⁻¹ sr⁻¹) |
|---|---|---|---|---|
| [0.95, 0.99] | < 0.3 | ≥ 105 | 0.18 | 7.27E-19 |
| [0.90, 0.95] | < 0.3 | ≥ 105 | 0.18 | 8.80E-19 |
| [0.86, 0.90] | < 0.3 | ≥ 105 | 0.18 | 1.24E-18 |
| [0.82, 0.86] | < 0.6 | ≥ 102 | 0.29 | 2.86E-18 |

**Table 1.** The optimized cuts, the number of background events remaining after cuts and the sensitivity obtained in each β range.

After applying the optimized cuts on the totality of the data taken, no event survived the selection. Figure 5 below shows the obtained ANTARES upper limit on the flux for Magnetic Monopoles, taking into account the full period of 2480 days of data taking, compared to the upper limits on the flux found by the other experiments such as IceCube, and including the upper limit (116 days in orange and 1012 days in red) of previous MM analyses carried out with the ANTARES telescope.

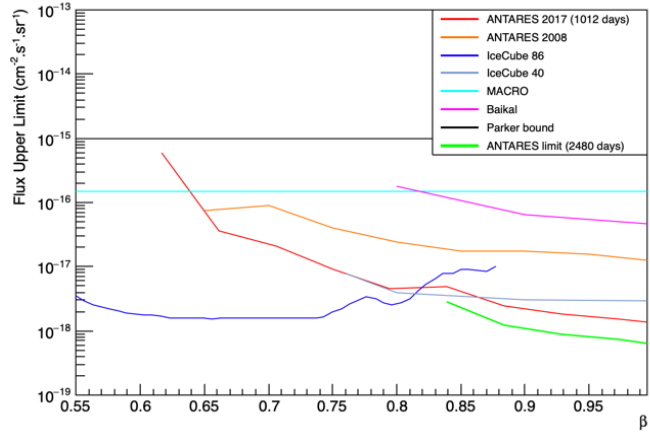

**Figure 5.** Antares upper limit on the flux of Magnetic Monopoles obtained (green line) compared to other experiments.

The limit for the Magnetic Monopole found in this analysis presents a good improvement compared with the last limit found by the Antares Detector.

No event survived the selection, the upper limits on the flux were set for each one of the 4 ranges of $β$ considered.